OXFORD

# Semiparametric sensitivity analysis: unmeasured confounding in observational studies

Razieh Nabi[1], Matteo Bonvini[2], Edward H. Kennedy[3], Ming-Yueh Huang[4], Marcela Smid[5], Daniel O. Scharfstein[6,*]

[1]Department of Biostatistics and Bioinformatics, Emory University, Atlanta, GA 30322, United States, [2]Department of Statistics, Rutgers University, Piscataway, NJ 08854, United States, [3]Department of Statistics & Data Science, Carnegie Mellon University, Pittsburgh, PA 15213, United States, [4]Institute of Statistical Science, Academia Sinica, Taipei 11529, Taiwan, [5]Department of Obstetrics and Gynecology, University of Utah School of Medicine, Salt Lake City, UT 84132, United States, [6]Department of Population Health Sciences, University of Utah School of Medicine, Salt Lake City, UT 84108, United States

[*]Corresponding author: Daniel O. Scharfstein, Department of Population Health Sciences, University of Utah School of Medicine, 295 Chipeta Way, Salt Lake City, Utah 84108, USA (daniel.scharfstein@hsc.utah.edu).

## ABSTRACT

Establishing cause–effect relationships from observational data often relies on untestable assumptions. It is crucial to know whether, and to what extent, the conclusions drawn from non-experimental studies are robust to potential unmeasured confounding. In this paper, we focus on the average causal effect (ACE) as our target of inference. We generalize the sensitivity analysis approach developed by Robins et al., Franks et al., and Zhou and Yao. We use semiparametric theory to derive the non-parametric efficient influence function of the ACE, for fixed sensitivity parameters. We use this influence function to construct a one-step, split sample, truncated estimator of the ACE. Our estimator depends on semiparametric models for the distribution of the observed data; importantly, these models do not impose any restrictions on the values of sensitivity analysis parameters. We establish sufficient conditions ensuring that our estimator has $\sqrt{n}$ asymptotics. We use our methodology to evaluate the causal effect of smoking during pregnancy on birth weight. We also evaluate the performance of estimation procedure in a simulation study.

**KEYWORDS:** causal inference; influence function; split sample estimation; truncation.

## 1 INTRODUCTION

In causal inference, we often seek to make inferences about the population effect of a binary treatment on an outcome variable by contrasting means of potential outcomes $Y(1)$ and $Y(0)$ (ie, counterfactuals), where $Y(t)$ represents the (bounded) outcome of a random individual under treatment $t$, (Neyman, 1923; Rubin, 1974). Identification of this contrast, called the average causal effect (ACE), from observational data (ie, non-experimental studies) requires untestable assumptions. Standard assumptions include:

1. *Consistency*: The observed outcome $Y$ is equal to the potential outcome $Y(t)$ when the treatment received is $t$, i.e., $T = t$;
2. *Conditional ignorability*: There exists a set of measured pre-treatment covariates $X$ such that treatment is conditionally independent of the potential outcomes given $X$, i.e.,
$$Y(t) \perp T \mid X \quad \text{for } t = 0, 1; \tag{1}$$
3. *Positivity*: For each level of the covariates $X$, the probability of receiving either treatment is greater than zero, i.e., $P(T = t \mid X = x) > 0$, for all $x$ in the state space of $X$.

Under these assumptions, the ACE is identified from the observed data distribution via the *adjustment formula*:

$$ACE = \int_x \left\{ \mathbb{E}[Y \mid T = 1, X = x] - \mathbb{E}[Y \mid T = 0, X = x] \right\} dF(x), \tag{2}$$

where $F(\cdot)$ denotes the cumulative distribution function of $X$. Using $n$ independent and identically distributed copies of $Z = (X, T, Y)$, many methods have been developed to draw inference about the ACE functional, e.g., propensity score matching (Rosenbaum and Rubin, 1983b), g-computation (Robins, 1986), (stabilized) inverse probability weighting (Hernán and Robins, 2006), augmented inverse probability weighting (Robins et al., 1994), and targeted maximum likelihood (van Der Laan and Rubin, 2006).

The conditional independence expressed in (1) implies that there are no unmeasured confounders between treatment and outcome. Assessing the robustness of inferences to potential unmeasured confounding is considered crucial. The goal of this manuscript is to provide a methodology for evaluating the sensitivity of inferences about the ACE to deviations from (1).

As motivation for our work, consider Almond et al. (2005) who sought to evaluate the causal effect of smoking during pregnancy on birth weight. Their analysis was based on approximately 500 000 singleton births in Pennsylvania between

Received: April 15, 2022; Revised: July 24, 2024; Accepted: September 22, 2024

1989 and 1991 and assumed no unmeasured confounding. After adjustment for a large collection of measured "pre-treatment variables" (maternal factors—demographic, prenatal care, pregnancy history, co-morbidities; paternal factors—demographic) in a regression model, they reported a reduction of 203.2 g in birth weight for smokers vs non-smokers. Important factors not controlled for in their analysis included maternal nutrition, social determinants of health, use of substances other than alcohol, and genetics and epigenetics. This raises the question: How does failure to control for potential unmeasured confounders impact inferences about the causal effect of smoking during pregnancy on birth weight? We will illustrate how to use the methodology that we develop in this paper to address this question.

In the next section, we provide a brief overview of the literature on sensitivity analysis in causal inference. In Section 3, we present a specific class of assumptions indexed by sensitivity analysis parameters that quantify departures from (1) and provide an identification formula for the ACE. In Section 4, we present our semiparametric strategy for drawing inference about the ACE under this class of assumptions. In Section 5, we reanalyze a subset of the Pennsylvania singleton birth dataset. Section 7 is devoted to a discussion.

## 2 PRIOR WORK ON SENSITIVITY ANALYSIS

Sensitivity analysis to the "no unmeasured confounders" assumption is designed to probe the impact of residual unmeasured confounding on causal effect estimates. One of the earliest works on sensitivity analysis is attributed to Cornfield et al. (1959). The work of Cornfield et al. (1959) is not immediately relevant for our setup as it focuses on relative risks and does not incorporate measured covariates. Recently, Ding and VanderWeele (2016) extended the approach of Cornfield et al. (1959) by allowing for low-dimensional measured covariates and introducing two sensitivity analysis parameters that govern the impact of unmeasured confounding on the outcome and treatment, respectively. They derive a bound on the relative risk in terms of the observed relative risk and the two sensitivity analysis parameters. Other sensitivity analysis methods have been developed for relative risks (see E-Appendix 4 of Ding and VanderWeele [2016]). In addition to focusing on relative risks, these methods do not accommodate complex measured confounders.

We now provide a more extensive review on sensitivity analysis for the ACE of a binary treatment. Our review divides approaches into two types: those that seek set identification and those that seek point identification of the ACE (at each sensitivity parameter value).

### 2.1 Set identification

#### 2.1.1 *Bounds without sensitivity parameters*

If no assumptions are made on the unmeasured confounders and the outcome is bounded, the ACE can be restricted to an interval informed by the observed distribution (Robins, 1989; Manski, 1990). The lower and upper bounds of this interval are computed under extreme instances of residual confounding. For this reason, the interval tends to be wide and necessarily includes zero. Robins (1989) and Manski (1990) derived tighter bounds by imposing additional non-identifiable assumptions.

#### 2.1.2 *Bounds with sensitivity parameters*

To achieve better control over departures from the no-unmeasured-confounding assumption, sensitivity analysis procedures have been proposed that bound the impact of an unmeasured confounder on the treatment and/or the outcome. One way is to assume that, for units sharing the same value of measured covariates but a different value of the unmeasured confounder, the odds ratio of the probabilities of receiving treatment differs by at most $\Gamma$. For matched studies, Rosenbaum (1987) developed a method for finding the minimum $\Gamma$ such that inference about the ACE is not "statistically significant". Gastwirth et al. (1998) extended this idea by additionally incorporating a bound on the impact of the unmeasured confounder on the outcome. Yadlowsky et al. (2018) extended the idea of Rosenbaum (1987) to general study designs. Shen et al. (2011) proposed bounds on ACE based on the variance of the multiplicative error in estimating the probability of receiving treatment given both measured and unmeasured confounding and the correlation between this error and the potential outcomes. Tan (2006) and Zhao et al. (2019) derived bounds on ACE by bounding the ratio of the odds of receiving treatment given measured and unmeasured confounders to the odds of receiving treatment given measured confounders. The bounds derived in Zhao et al. (2019) have been recently improved by Dorn and Guo (2022) and Dorn et al. (2021). The resulting bounds have a closed-form expression that depends on the observed propensity score, a certain transformed-outcome regression and the conditional quantiles of the outcome given the treatment and the covariates. Furthermore, Dorn et al. (2021) showed that it is possible to construct estimators of these bounds that remain valid, albeit conservative, even if the conditional quantiles are misspecified as long as at least one of the other two nuisance functions is consistently estimated.

In other work, Díaz and van der Laan (2013) and Díaz et al. (2018) derived bounds on the ACE by bounding the difference of the mean potential outcome had patients received treatment or control, given covariates, among those who actually received treatment versus control. Bonvini and Kennedy (2022) took a contamination model approach, giving bounds on the ACE by constraining the proportion of units affected by unmeasured confounding.

### 2.2 Point identification

An alternative approach is to posit sensitivity analysis parameters that admit identification of the ACE. A number of authors have proposed using sensitivity analysis parameters to govern the relationship among unmeasured confounder(s), outcome and treatment. Rosenbaum and Rubin (1983a) developed a methodology that handles low-dimensional measured covariates, binary treatment, binary outcome, and a binary unmeasured confounder. This approach has been extended to accommodate normally distributed outcomes (Imbens, 2003), continuous treatments and a normally distributed unmeasured confounder (Carnegie et al., 2016), and a semiparametric Bayesian approach when the treatment and unmeasured confounder are binary (Dorie et al., 2016).

In order to avoid positing a marginal distribution for the unmeasured confounder(s), Zhang and Tchetgen Tchetgen

(2022) devised a semiparametric approach to sensitivity analysis that requires models for the conditional probability of receiving treatment and the conditional mean of the outcomes; a sensitivity analysis parameter in each of these models governs the influence of the unmeasured confounder. Veitch and Zaveri (2020) circumvented the need to model the marginal distribution of the unmeasured confounders by specifying a propensity score model depending on measured and unmeasured covariates; the model is anchored at a propensity score model that is only conditional on measured confounders and is indexed by a sensitivity analysis parameter governing the influence of the unmeasured confounder. They also introduced a sensitivity analysis parameter that governs the influence of the propensity score on the conditional mean of outcome given treatment and measured covariates.

VanderWeele and Arah (2011) introduced a general "bias" formula for the difference between the possibly incorrect expression for the ACE under no unmeasured confounding and the correct expression for the ACE when accounting for both measured and unmeasured confounding in terms of many sensitivity parameters. They introduced simplifying assumptions in order to require fewer sensitivity parameters in their formalization and make it easier to use in practice. Cinelli and Hazlett (2020) focused on the linear model setting and introduced an "omitted-variable" bias formula that depends on partial $R^2$ values that govern the association between the unmeasured confounder, the outcome, and the treatment; these $R^2$ values are specified as sensitivity analysis parameters.

Brumback et al. (2004) and Robins (1999) discussed a sensitivity analysis methodology for unmeasured confounding in the setting of the time-varying treatment regimens. In the context of a point exposure, their approach and that of Sjölander et al. (2022) and Lu and Ding (2023) is tantamount to specifying a sensitivity analysis function that governs the difference in the conditional (on measured covariates) means of the outcome under treatment (control) between treated and untreated individuals. Robins et al. (2000), Franks et al. (2020), and Zhou and Yao (2023) specified sensitivity parameters that govern a contrast between the conditional (on measured covariates) distributions of the outcome under treatment (control) between treated and untreated individuals. An attractive feature of these approaches is that the sensitivity analysis specification does not impose any restrictions on the distribution of the observed data while yielding identification of the ACE. In this paper, we generalize the work of Robins et al. (2000), Franks et al. (2020), and Zhou and Yao (2023) and propose an estimator for the ACE using semiparametric efficiency theory.

## 3 SENSITIVITY ANALYSIS MODEL AND IDENTIFICATION

### 3.1 Notation

Let the covariate set $X$ be of dimension $p$. We let $P$ denote the true distribution of the observed data, which is characterized by $F(x) = P(X \le x)$, $\pi_t(x) = P(T = t \mid X = x)$, and $F_t(y \mid x) = P(Y \le y \mid T = t, X = x)$. Let $\mu_t(g(Y, x); x) = \mathbb{E}[g(Y, x) \mid T = t, X = x] = \int g(y, x) dF_t(y \mid x)$.

When we refer to an estimator of $P$, denoted by $\widehat{P}$, we are referring to $\widehat{\pi}_t(x)$, $\widehat{F}_t(y|x)$ (including associated functionals), and $\widehat{F}(x)$ (taken to be the empirical distribution of $X$). We let $P_n$ denote the empirical distribution of the observed data based on $n$ observations. In our estimation procedure, we will randomly split the observations into $K$ disjoint sets, where $S_i$ denotes the split membership of the $i$th observation (ie, $S_i \in \{1, \dots, K\}$). The size of the $k$th disjoint set is denoted by $n_k \approx n/K$ (ie, $n_k = O(n)$). We let $\widehat{P}^{(-k)}$ be an estimator of $P$ based on observations from all the splits except that of $k$th split. We let $P_{n_k}^{(k)}$ be the empirical distribution of the observed data in the $k$th split.

Throughout, $||f(Z)||_{L_2} := \sqrt{\int f(z)^2 \, dP(z)}$.

### 3.2 Model

We seek to identify $\mathbb{E}[Y(t)]$ by positing untestable assumptions that identify the non-identified distribution of $Y(t)$ given $T = 1 - t$ and $X = x$. We accomplish this by positing a model that connects this distribution to the identifiable distribution of $Y(t)$ given $T = t$ and $X = x$. Specifically, we consider a model of form:

$$dF(y(t) \mid T = 1 - t, X = x) = \underbrace{dF(y(t) \mid T = t, X = x)}_{dF_t(y|x)} q_t\{y, x; \gamma_t, h_t(F_t(\cdot|x); \gamma_t)\}, \quad (3)$$

where $h_t(\cdot; \gamma_t)$ is a scalar functional for each $x$, $q_t\{y, x; \gamma_t, h_t(F_t(\cdot|x); \gamma_t)\}$ is a specified non-negative function such that (1) $q_t\{y, x; \gamma_t, h_t(F_t(\cdot|x); \gamma_t)\}$ equals 1 only when $\gamma_t = 0$ and (2) $\int dF_t(y|x) q_t\{y, x; \gamma_t, h_t(F_t(\cdot|x); \gamma_t)\} = 1$ for all $x$ and $\gamma_t$. Here, $\gamma_t$ is a sensitivity parameter and $h_t(\cdot; \gamma_t)$ is introduced to ensure that $dF(y(t)|T = 1 - t, X = x)$ integrates to 1. Note that the choice of $\gamma_t = 0$ corresponds to the no unmeasured confounding assumption. An important feature of Assumption (3) is that it places no restrictions on $P$.

Robins et al. (2000), Franks et al. (2020), and Zhou and Yao (2023) considered special cases of (3). Specifically, Robins et al. (2000) assumed

$$q_t\{y, x; \gamma_t, h_t(F_t(\cdot|x))\} = \frac{\exp\{s_t(y, x; \gamma_t)\}}{\mu_t(\exp\{s_t(Y, x; \gamma_t)\}; x)}, \quad (4)$$

where $s_t(y, x; \gamma_t)$ is a specified, bounded function of its arguments with $s_t(y, x; \gamma_t) = 0$ only when $\gamma_t = 0$. Here, $h_t(F_t(\cdot|x); \gamma_t) = \mu_t(\exp\{s_t(Y, x; \gamma_t); x)$. Franks et al. (2020) assumed that $s_t(y, x; \gamma_t) = \gamma_t s_t(y)$, where $s_t(y)$ is a specified function of $y$. Zhou and Yao (2023) considered the further special case where $s_t(y) = y$.

Model (4) has the following features:

1. When $s_t(y, x; \gamma_t)$ is increasing (decreasing) in $y$, the conditional (on $X$) density of $Y(t)$ for individuals with $T = 1 - t$ is shifted toward higher (lower) values relative to the conditional density for individuals with $T = t$.
2. When the derivative of $s_t(y, x; \gamma_t)$ with respect to $y$ changes from positive (negative) to negative (positive), the conditional (on $X$) density of $Y(t)$ for individuals with $T = 1 - t$ is shifted toward extreme (central) values relative to the conditional density for individuals with $T = t$.

Model (4) is that it has difficulty aligning the tails of $dF(y(t)|T = 1 - t, X = x)$ with the tails of $dF(y(t) \mid T = t, X = x)$ over a range of $\gamma_t$. This is because the normalization factor in (4) does not depend on $y$. The tail restriction is important because in our motivating application, it is thought that extremely low birth weights (ELBW) (<1000 g) and extreme macrosomia (>5000 g) are primarily due to factors other than smoking. For ELBW, other significant risk factors include maternal infection (Palmsten et al., 2018), pregnancy-induced hypertension (Salafia et al., 1995) and placental pathologies (Salafia et al., 1995). Risk factors for extreme macrosomia include pregestational and gestational diabetes (Berard et al., 1998) and advanced gestational age (Stotland et al., 2004).

Abstractly, we would like to impose a substantive restriction that $dF(y(t)|T = 1 - t, X) \approx dF(y(t)|T = t, X)$ for all $y(t)$ less than some threshold $c_l(t)$ and greater than some threshold $c_u(t)$. To accomplish this, we consider a special case of (3) where

$$
\begin{aligned}
&q_t\{y, x; \gamma_t, h_t(F_t(\cdot|x); \gamma_t)\} \\
&= \underbrace{1 - \mathcal{L}\left\{\frac{y - c_l(t)}{\epsilon(t)}\right\}}_{\approx I\{y \le c_l(t)\}} + \underbrace{\mathcal{L}\left\{\frac{y - c_u(t)}{\epsilon(t)}\right\}}_{\approx I\{y \ge c_u(t)\}} \\
&+ \underbrace{\left(\mathcal{L}\left\{\frac{y - c_l(t)}{\epsilon(t)}\right\} - \mathcal{L}\left\{\frac{y - c_u(t)}{\epsilon(t)}\right\}\right)}_{\approx I(c_l(t) \le y \le c_u(t))} \\
&\exp\{\gamma_t u_t(y, \mathbf{c}(t))\} h_t(F_t(\cdot|x); \gamma_t) \qquad (5)
\end{aligned}
$$

$\mathbf{c}(t) = (c_l(t), c_u(t))$, $\epsilon(t)$ is a small positive constant, $u_t(y, \mathbf{c}(t))$ is a specified continuous, bounded, function of $y$ and $\mathbf{c}(t)$, $u_t(c_l(t), \mathbf{c}(t)) = u_t(c_u(t), \mathbf{c}(t)) = 0$, $\mathcal{L}\{\cdot\}$ is a cumulative distribution function of a continuous random variable with symmetric density function around zero, and

$$
\begin{aligned}
&h_t(F_t(\cdot|x); \gamma_t) \\
&= \frac{\mu_t\left(\mathcal{L}\left\{\frac{y - c_l(t)}{\epsilon(t)}\right\} - \mathcal{L}\left\{\frac{y - c_u(t)}{\epsilon(t)}\right\}; x\right)}{\mu_t\left(\left(\mathcal{L}\left\{\frac{y - c_l(t)}{\epsilon(t)}\right\} - \mathcal{L}\left\{\frac{y - c_u(t)}{\epsilon(t)}\right\}\right)\exp\{\gamma_t u_t(Y, \mathbf{c}(t))\}; x\right)}. \qquad (6)
\end{aligned}
$$

In our motivating application, there is a belief expressed by our clinical co-author (MS) that, within levels of measured covariates,

1. The central part of the distribution of birth weight under smoking, for non-smokers, will be concentrated at "better" values than the central part of the distribution of birth weight under smoking, for smokers; here "better" means closer to 3400 g.
2. The central part of the distribution of birth weight under not smoking, for smokers, will be concentrated at lower values than the central part of the distribution of birth weight under not smoking, for non-smokers.

To incorporate these beliefs, we chose

$$
\begin{aligned}
&u_t(y, \mathbf{c}(t)) \\
&= \underbrace{\left(1 - \mathcal{L}\left\{\frac{y - c^*(t)}{\epsilon(t)}\right\}\right)}_{\approx I\{y \le c^*(t)\}} \frac{c^*(t)}{c^*(t) - c_l(t)}\{y - c_u(t)\} \\
&+ \underbrace{\mathcal{L}\left\{\frac{y - c^*(t)}{\epsilon(t)}\right\}}_{\approx I\{y \ge c^*(t)\}} \frac{c^*(t)}{c^*(t) - c_u(t)}\{y - c_l(t)\} \quad c_l(t) \le y \le c_u(t), \qquad (7)
\end{aligned}
$$

where $c_l(t) < c^*(t) < c_u(t)$; this smoothed triangular function increases from 0 to $c^*(t)$ and then decreases from $c^*(t)$ to 0. In our analysis, we set $\mathcal{L}(\cdot)$ to be cumulative distribution of function of a standard normal random variable, $\epsilon(1) = \epsilon(0) = 500$, $c_l(1) = c_l(0) = 1000$, $c_u(1) = c_u(0) = 5000$, $c^*(1) = 3400$ and $c^*(0) = 2000$. We considered values of $\gamma_1$ and $\gamma_0$ ranging from 0.0 to 0.0025. In top (bottom) panel of Figure 1, we display the distribution of birth weight under smoking (non-smoking) for non-smokers (smokers), for various choices of sensitivity analysis parameters, relative to estimated density of observed birth weights for smokers (non-smokers). Notice how the tails match, the distributions vary smoothly with $\gamma_t$, and there are substantial differences between the distributions when $\gamma_t = 0.0$ vs $\gamma_t = 0.0025$.

### 3.3 Identification

For fixed $\gamma_t$, it can be shown that $\mathbb{E}[Y(t)]$ is identified via the following formula:

$$
\begin{aligned}
&\psi_t(P; \gamma_t) \\
&= \int_x \left\{\mu_t(Y; x)\pi_t(x) + \mu_t(Y q_t\{y(t), X; \gamma_t, h_t(F_t(\cdot|X); \gamma_t)\}; x)\pi_{1-t}(x)\right\} dF(x). \qquad (8)
\end{aligned}
$$

Then, the ACE is identified by $\psi_1(P; \gamma_1) - \psi_0(P; \gamma_0)$. Note that $\psi_1(P; 0) - \psi_0(P; 0)$ reduces to (2). Since Assumption (3) places no restrictions on $P$, there can only be one non-parametric influence function for $\mathbb{E}[Y(t)]$ and *ACE* (Tsiatis, 2006).

## 4 INFERENCE

We derive semiparametric estimators for $\mathbb{E}[Y(t)]$ for $t = 0, 1$ and then combine the resulting estimators to draw inference about the ACE. We start by first deriving the influence function for $\mathbb{E}[Y(t)]$ under (3). The influence function can be used to "debias" plug-in estimators. In this context, the influence function of $\mathbb{E}[Y(t)]$, denoted $\phi_t(P; \gamma_t)(Z)$, depends on the distribution $P$; the negative of its expectation with respect to $P$ when evaluated at $\widehat{P}$, $-\int \phi(\widehat{P}; \gamma_t)(z)dP(z)$, expresses the first-order bias of the plug-in estimator $\psi_t(\widehat{P}; \gamma_t)$. We can thus construct a "debiased plug-in" (ie, "one-step") estimator by adding an estimate of $\int \phi(\widehat{P}; \gamma_t)(z)dP(z)$ to $\psi_t(\widehat{P}; \gamma_t)$. We will see that this influence function depends on $\pi_t(X)$ and $F_t(\cdot|X)$, which we propose to model semiparametrically. We then propose a one-step sample-split estimator for $\mathbb{E}[Y(t)]$. Kennedy (2022) provides a recent review of this approach.

**Theorem 1** ***Non-Parametric Influence Function***

$$
\begin{aligned}
\phi_t(P;\gamma_t)(Z) &= I(T=t)Y \\
&+ I(T=t)\left\{\frac{\pi_{1-t}(X)}{\pi_t(X)} \times \left(Yq_t\{Y,X;\gamma_t,h_t(F_t(\cdot|X);\gamma_t)\} - \mu_t(Yq_t\{Y,X;\gamma_t,h_t(F_t(\cdot|X);\gamma_t)\};X)\right)\right\} \\
&- I(T=t)\left\{\frac{\pi_{1-t}(X)}{\pi_t(X)} \times \frac{\mu_t(Yq_t'\{Y,X;\gamma_t,h_t(F_t(\cdot|X);\gamma_t)\};X)}{\mu_t(q_t'\{Y,X;\gamma_t,h_t(F_t(\cdot|X);\gamma_t)\};X)}\right. \\
&\qquad \left. \times \left(q_t\{Y,X;\gamma_t,h_t(F_t(\cdot|X);\gamma_t)\} - \mu_t(q_t\{Y,X;\gamma_t,h_t(F_t(\cdot|X);\gamma_t)\};X)\right)\right\} \\
&+ I(T=1-t)\mu_t(Yq_t\{Y,X;\gamma_t,h_t(F_t(\cdot|X);\gamma_t)\};X) - \psi_t(P;\gamma_t),
\end{aligned} \tag{9}
$$

*where $q_t'\{y,x;\gamma_t,h_t(F_t(\cdot|x);\gamma_t)\}$ is the derivative of $q_t$ with respect to the $h_t$ argument.*

**Proof.** See Supplementary Appendix A. □

Under (4), the non-parametric influence function for $\mathbb{E}[Y(t)]$ reduces to

$$
\begin{aligned}
\phi_t(P;\gamma_t)(Z) = I(T=t)Y + \\
I(T=t)\frac{\pi_{1-t}(X)}{\pi_t(X)}\frac{\exp\{s_t(Y,X;\gamma_t)\}}{\mu_t(\exp\{s_t(Y,X;\gamma_t)\};X)}\left\{Y - \frac{\mu_t(Y\exp\{s_t(Y,X;\gamma_t)\};X)}{\mu_t(\exp\{s_t(Y,X;\gamma_t)\};X)}\right\} \\
+ (T=1-t)\frac{\mu_t(Y\exp\{s_t(Y,X;\gamma_t)\};X)}{\mu_t(\exp\{s_t(Y,X;\gamma_t)\};X)} - \psi_t(P;\gamma_t).
\end{aligned} \tag{10}
$$

**Remark 1** *There is a connection between the influence function (10) and one that has been derived in the missing data literature. Let $R_t = I(T=t)$ be the missing data indicator associated with observation of $Y(t)$, i.e., $Y(t)$ is observed when $R_t = 1$ and unobserved when $R_t = 0$. With this notation, Assumption (4) is identical to that considered in the missing data literature (see, eg, Scharfstein et al. [1999] and Scharfstein and Irizarry [2003]). This assumption places no restrictions on the law $(P_t)$ of $Z_t = (X, R_t, Y(t) : R_t = 1)$ or the law $(P)$ of $Z = (X, T, Y)$; note that $P_t$ can be derived from $P$. The identifying functional for $\mathbb{E}[Y(t)]$ expressed in (8) depends on $P$ through $P_t$. Thus, the non-parametric influence function for this functional based on data $Z_t$, $\phi_t(P_t;\gamma_t)(Z_t)$, is an influence function for $\mathbb{E}[Y(t)]$ based on data $Z$. Since the non-parametric influence function for $\mathbb{E}[Y(t)]$ based on $Z$ must be unique under Assumption (4), it must be equal to $\phi_t(P_t;\gamma_t)(Z_t)$, which has been previously derived.*

In Supplementary Appendix B, we derive a formula for $\mathrm{Rem}_t(\widetilde{P},P;\gamma_t) := \psi_t(\widetilde{P};\gamma_t) - \psi_t(P;\gamma_t) + \mathbb{E}[\phi_t(\widetilde{P};\gamma_t)(Z)]$, where $\widetilde{P}$ is any distribution of the observed data. $\mathrm{Rem}_t(\widehat{P},P;\gamma_t)$ is the residual bias of the one-step estimator and it can be seen to be an expectation of a product of differences (ie, second-order). We use this formula to show why we can estimate $F_t(y|X)$ and $\pi_t(X)$ at rates slower than $\sqrt{n}$ and still obtain a $\sqrt{n}$-rate for our estimator of $\mathbb{E}[Y(t)]$.

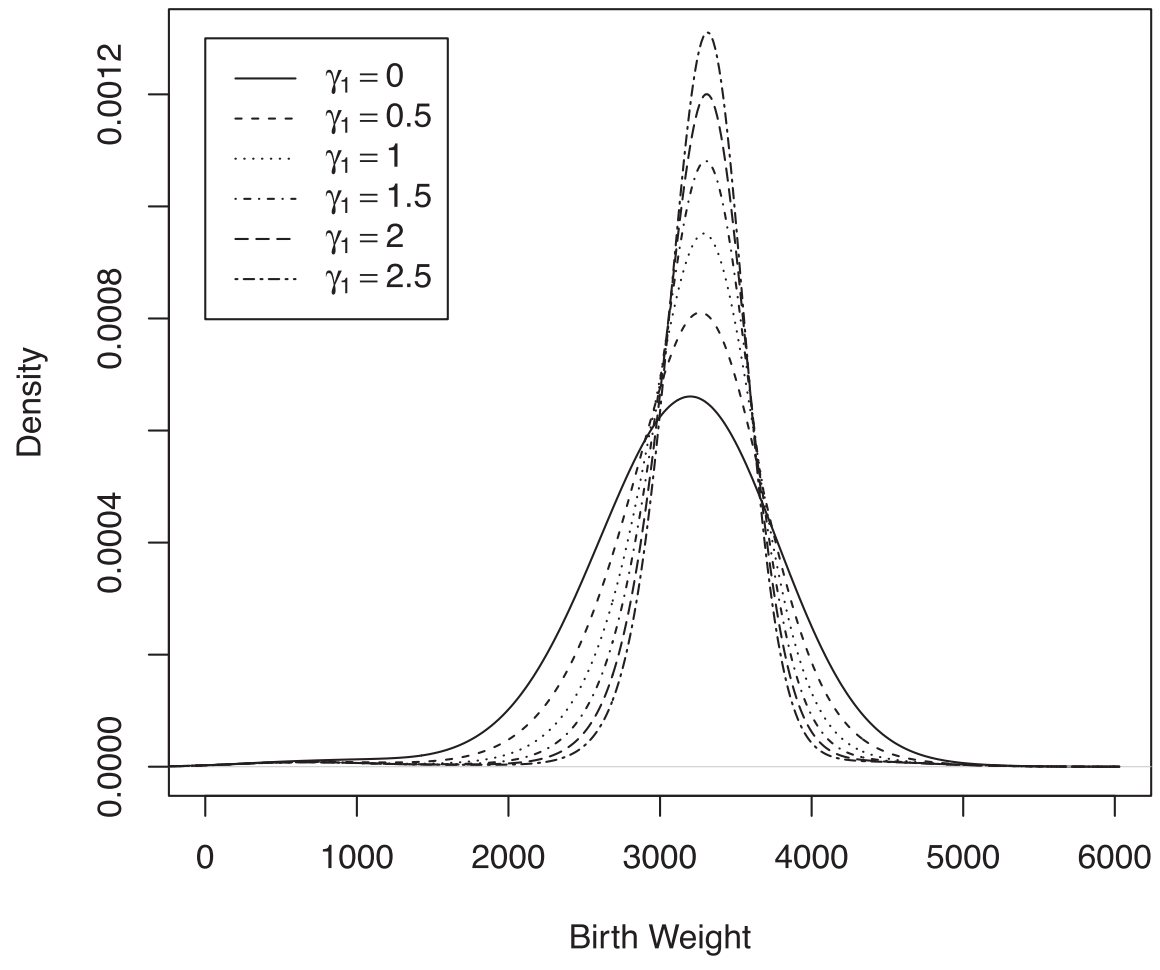

(a) Smokers

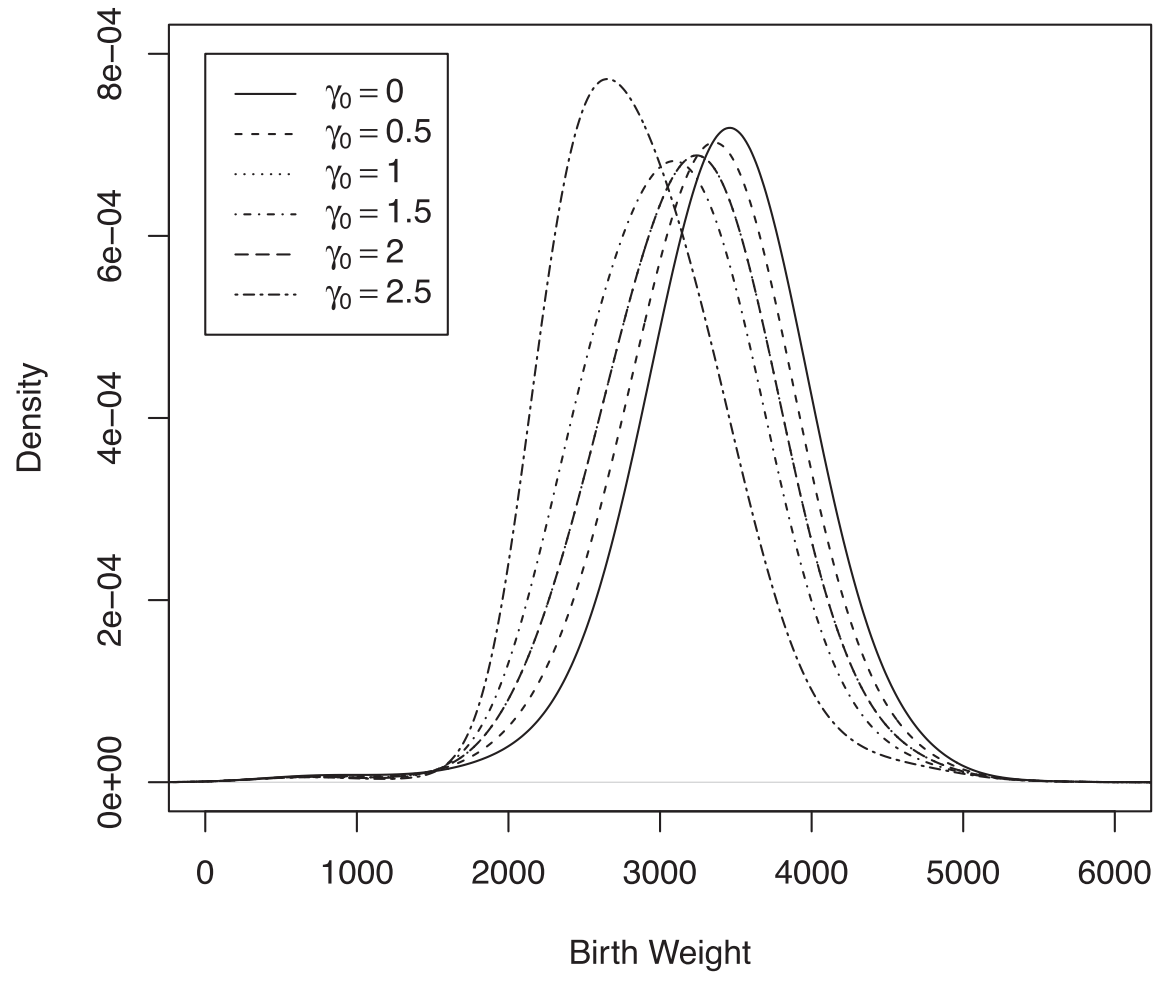

(b) Non-smokers

**FIGURE 1** Distribution of birth weight (a) under smoking for non-smokers, and (b) under non-smoking for smokers, for various sensitivity analysis parameters ($\times 1000$), relative to the estimated densities of observed birth weights for the respective groups.

### 4.1 Modeling and estimation of *P*

In Model (4), Rotnitzky et al. (2021) suggested modeling $r_{1t}(X;\gamma_t) = \frac{\pi_{1-t}(X)}{\pi_t(X)\mu_t(\exp\{s_t(Y,X;\gamma_t)\};X)}$ and $r_{2t}(X;\gamma_t) =$ $\frac{\mu_t(Y\exp\{s_t(Y,X;\gamma_t)\};X)}{\mu_t(\exp\{s_t(Y,X;\gamma_t)\};X)}$. The advantage of this approach is that, for fixed $\gamma_t$, one can obtain estimators of $\psi_t(P;\gamma_t)$ that are consistent and asymptotically normal if either the model for $r_{1t}(X;\gamma_t)$ or the model for $r_{2t}(X;\gamma_t)$ is correctly specified. It is unclear how to abstract this modeling approach to the general form of Model (3). Nevertheless, a key disadvantage of this approach is that modeling and associated model evaluation is *entangled* with $\gamma_t$. The approach we adopt is to flexibly model $\pi_t(X)$ and $F_t(\cdot|X)$. While this approach is robust to mis-specification of $\pi_t(X)$, it does rely on correct specification of a model for $F_t(\cdot|X)$. An important advantage is that model evaluation does not depend on $\gamma_t$.

We posit a generalized additive model (GAM) with a logistic link function for $\pi_t(X)$ (Hastie and Tibshirani, 2017). Let $\widehat{\pi}_t(X)$ be the GAM estimator of $\pi_t(X)$. Horowitz and Mammen (2004) showed that $\left|\left|\widehat{\pi}_t(X)-\pi_t(X)\right|\right|_{L_2}=O_P(n^{-2/5})$. We posit a single index model (Chiang and Huang, 2012) for $F_t(\cdot|X)$. This model assumes that $F_t(y|X)=F_t(y,X'\beta_t;\beta_t)$, where $F_t(y,u;\beta_t)$ is a cumulative distribution function in $y$ for each $u$, $\beta_t=(\beta_{1,t},\ldots,\beta_{p,t})$ is a vector of unknown parameters and, for purposes of identifiability, $\beta_{1,t}$ is set to 1. We estimate $F_t(y,x'\beta_t;\beta_t)$ by

$$\widehat{F}_t(y,X'\widehat{\beta}_t;\widehat{\beta}_t)=\frac{\sum_{T_i=t}I(Y_i\le y)\mathcal{K}_{\widehat{b}_t}(X_i'\widehat{\beta}_t-X'\widehat{\beta}_t)}{\sum_{T_i=t}\mathcal{K}_{\widehat{b}_t}(X_i'\widehat{\beta}_t-X'\widehat{\beta}_t))},$$

where $\mathcal{K}_b(v)=\mathcal{K}(v/b)/b$, $\mathcal{K}$ is a $r$th-order kernel, $\widehat{\beta}_t$ is an estimator of $\beta_t$ and $\widehat{b}_t$ is an estimator of the bandwidth $b$ (see Supplementary Appendix C for estimation details).

To simplify the analysis of our estimator, we assume that $h_t(F_t(\cdot|X);\gamma_t)$ itself can be written as a differentiable function (with uniformly bounded gradient) of a fixed number of regression functions of the form $\mu_t(\rho_{j,t}(Y,X;\gamma_t);X)$, $j\in\{1,\ldots,J_t\}$, $J_t<\infty$, for some known bounded outcomes $\rho_{j,t}(Y,X;\gamma_t)$. This is the case for all the choices of $q_t(\cdot)$ described in this work. For example, when $q_t$ is specified as in (5), $J_t=2$, $\rho_{1,t}(Y,X;\gamma_t)=\mathcal{L}\left\{\frac{y-c_l(t)}{\epsilon(t)}\right\}-\mathcal{L}\left\{\frac{y-c_u(t)}{\epsilon(t)}\right\}$ and $\rho_{2,t}(Y,X;\gamma_t)=\left(\mathcal{L}\left\{\frac{y-c_l(t)}{\epsilon(t)}\right\}-\mathcal{L}\left\{\frac{y-c_u(t)}{\epsilon(t)}\right\}\right)\exp\{\gamma_t u_t(Y,\mathbf{c}(t))\}$. In order to use the non-parametric influence function (9) for estimation of $\psi_t(P;\gamma_t)$, we first estimate $h_t(F_t(\cdot|X);\gamma_t)$ by $h_t(\widehat{F}_t(\cdot|X);\gamma_t)$ which is computed by plugging in estimators of $\mu_t(\rho_{j,t}(Y,X;\gamma_t);X)$, $j=1,\ldots,J_t$. Specifically, for a general function $\rho(Y,X;\gamma_t)$, we estimate $\mu_t(\rho(Y,X;\gamma_t);X)$ by

$$\widehat{\mu}_t(\rho(Y,X;\gamma_t);X)=\int\rho(y,X;\gamma_t)d\widehat{F}_t(y,X'\widehat{\beta}_t;\widehat{\beta}_t). \quad (11)$$

We then estimate $\mu_t\{Yq_t\{Y,X;\gamma_t,h_t(F_t(\cdot|X);\gamma_t);X\}$, $\mu_t\{q_t\{Y,X;\gamma_t,h_t(F_t(\cdot|X);\gamma_t);X\}$, $\mu_t\{Yq_t'\{Y,X;\gamma_t,h_t(F_t(\cdot|X);\gamma_t);X\}$, and $\mu_t\{q_t'\{Y,X;\gamma_t,h_t(F_t(\cdot|X);\gamma_t)\};X\}$ in (9) by using (11) to, respectively, compute $\widehat{\mu}_t\{Yq_t\{Y,X;\gamma_t,h_t(\widehat{F}_t(\cdot|X);\gamma_t);X\}$, $\widehat{\mu}_t\{q_t\{Y,X;\gamma_t,h_t(\widehat{F}_t(\cdot|X);\gamma_t);X\}$, $\widehat{\mu}_t\{Yq_t'\{Y,X;\gamma_t,h_t(\widehat{F}_t(\cdot|X);\gamma_t);X\}$, and $\widehat{\mu}_t\{q_t'\{Y,X;\gamma_t,h_t(\widehat{F}_t(\cdot|X);\gamma_t)\};X\}$. Lastly, we estimate $\pi_t(X)$ in (9) by $\widehat{\pi}_t(X)$. We refer to the collection of functions $\mu_t(\cdot|X)$ and $\pi_t(X)$ as "nuisance functions," i.e., functions of the distribution $P$ that are not the estimand of interest but nevertheless need to be estimated.

In Supplementary Appendix C, we establish conditions under which, for a general function $\rho(Y,X;\gamma_t)$,

$$\sup_x\left|\widehat{\mu}_t(\rho(Y,X;\gamma_t);x)-\mu_t(\rho(Y,X;\gamma_t);x)\right|$$
$$=O_P\left(\left(\frac{\log n}{n}\right)^{\frac{r}{2r+1}}\right), \quad (12)$$

where $r$ is such that $P(Y\le y|T=t,X'\beta=u)$ and the density of $X'\beta$ have Lipschitz $(r+1)$th-order derivatives. In Theorem 2, we establish the $\sqrt{n}$-consistency and asymptotic normality of our estimators under the assumption that the nuisance functions' errors are asymptotically negligible. From our proof, a sufficient condition ensuring such negligibility is that, for each $\mu_t(\cdot\mid X)$, its squared estimation error as well as its estimation error times the error in estimating $\pi_t(X)$ are $o_P(n^{-1/2})$, where the errors are measured in $L_2$ norm. Because $\|\widehat{\pi}_t(X)-\pi_t(X)\|_{L_2}=O_P(n^{-2/5})$, it can be seen that $r\ge 1$ in (12) is a sufficient condition.

### 4.2 Estimation of $\mathbb{E}[Y(t)]$

The one-step, split sample estimator (Kennedy, 2022) of $\mathbb{E}[Y(t)]$ takes the form:

$$\widehat{\psi}_t(\gamma_t)$$
$$=\frac{1}{K}\sum_{k=1}^{K}\underbrace{\left\{\psi_t\left(\widehat{P}^{(-k)};\gamma_t\right)+\frac{1}{n_k}\sum_{i:S_i=k}\phi_t\left(\widehat{P}^{(-k)};\gamma_t\right)(Z_i)\right\}}_{k\text{th Split One-Step Estimator }\left(\widehat{\psi}_t^{(k)}(\gamma_t)\right)}. \quad (13)$$

Given the form of $\phi_t(P;\gamma_t)(Z)$ as an augmented inverse weighted influence function, there can be some numerical instability due to small predicted probabilities of treatment received. To address this issue, we use the tuning-free Huberization procedure developed by Wang et al. (2021). In this approach, the $k$th split estimator is replaced by

$$\widetilde{\psi}_t^{(k)}(\gamma_t)$$
$$=\frac{1}{n_k}\sum_{i:S_i=k}\underbrace{\min\left\{|\nu_t(\hat{P}^{(-k)};\gamma_t)(Z_i)|,\widehat{\tau}_t^{(k)}\right\}\operatorname{sign}\left\{\nu_t(\hat{P}^{(-k)};\gamma_t)(Z_i)\right\}}_{\widetilde{\nu}_t(\hat{P}^{(-k)};\gamma_t)(Z_i)},$$

where $\widehat{\tau}_t^{(k)}$ is the non-negative solution to

$$\frac{1}{n_k}\sum_{i:S_i=k}\frac{\min\left\{\nu_t(\hat{P}^{(-k)};\gamma_t)(Z_i)^2,\left(\tau_t^{(k)}\right)^2\right\}}{\left(\tau_t^{(k)}\right)^2}=\frac{\log(n_k)}{n_k}$$

and $\nu_t(\hat{P}^{(-k)};\gamma_t)(Z_i)=\psi_t\left(\widehat{P}^{(-k)};\gamma_t\right)+\phi_t\left(\widehat{P}^{(-k)};\gamma_t\right)(Z_i)$. This procedure effectively truncates $\nu_t(\hat{P}^{(-k)};\gamma_t)(Z_i)$ to be in $[-\tau_t^{(k)},\tau_t^{(k)}]$. We denote the one-step, split sample truncated estimator by $\widetilde{\psi}_t(\gamma_t)=\frac{1}{K}\sum_{k=1}^{K}\widetilde{\psi}_t^{(k)}(\gamma_t)$.

In Supplementary Appendix C, we prove the following theorem:

**Theorem 2** *Assume (1) $|Y|$ has compact support, (2) the function $u \mapsto q_t(y, x; \gamma_t, u)$ is twice differentiable with bounded second derivative for all $(y, x, \gamma_t)$ and (3) $0 < \pi_t(X) < 1$ and $\mu_t(q_t'\{Y, X; \gamma_t, h_t(F_t(\cdot|X); \gamma_t); X) \neq 0$ with probability one. Under Model (3) and the modeling assumptions/estimation approach described in Section 4.1,*

$$\sqrt{n}\left\{\widetilde{\psi}_t(\gamma_t) - \psi_t(P;\gamma_t)\right\} \overset{D(P)}{\to} \mathcal{N}(0, \mathbb{E}[\phi_t(P;\gamma_t)(Z)^2])$$

$$\sqrt{n}\left\{(\widetilde{\psi}_1(\gamma_1) - \widetilde{\psi}_0(\gamma_0)) - (\psi_1(P;\gamma_1) - \psi_0(P;\gamma_0)\right\}$$

$$\overset{D(P)}{\to} \mathcal{N}(0, \mathbb{E}[\{\phi_1(P;\gamma_1)(Z) - \phi_0(P;\gamma_0)(Z)\}^2]).$$

A central element in the proof of $\sqrt{n}$-consistency of our estimator is the requirement that $\sqrt{n}Rem_t(\widehat{P}, P; \gamma_t)$ is asymptotically negligible, which is satisfied under the modeling assumptions/estimation approach in Section 4.1.

We estimate the variance of $\widetilde{\psi}_t(\gamma_t)$ by $\frac{1}{nK}\sum_{k=1}^{K}\widehat{\sigma}_k^2(\gamma_t)$, where

$$\widehat{\sigma}_k^2(\gamma_t) = \frac{1}{n_k - 1}\sum_{i:S_i=k}\{\widetilde{\nu}_t(\hat{P}^{(-k)};\gamma_t)(Z_i) - \widetilde{\psi}_t^{(k)}(\gamma_t)\}^2.$$

This variance can be used to construct Wald-based confidence intervals.

## 5 DATA ANALYSIS

We analyzed a sample of 4996 singleton births in Pennsylvania between 1989 and 1991. The average birth weight is 3133 g and 3412 g for smokers and non-smokers, respectively. The naive estimated difference (smokers minus non-smokers) is −278 g (95% CI: −319 to −238). In our analysis, we accounted for the following maternal covariates: (1) age, (2) education (less than high school, high school, greater than high school), (3) white (yes/no), (4) hispanic (yes/no), (5) foreign (yes/no), (6) alcohol use, (7) married (yes/no), (8) liver birth order (one, two, and greater than 2), and (9) number of prenatal visits. Table 1 shows summary statistics for these covariates, stratified by maternal smoking status. There are striking differences between smokers and non-smokers, especially with respect to education, marital status, alcohol use and live birth order.

To evaluate goodness of fit, we simulated two large datasets, each comprised of observed data on 100 000 inidviduals. Both datasets were generated using the empirical distribution of covariates. The first dataset ("semiparametric") is generated using the estimated fits of the GAM and single-index models. The second dataset ("parametric") is generated using the estimated fits of a logistic regression model for the probability of $T = 1$ given $X$ and normal regression models for conditional distribution of $Y$ given $T = t$ and $X$. We computed empirical distributions of birth weight for smokers and non-smokers based on the original ("nonparametric") dataset, the semiparametric dataset and the parametric dataset. We then computed Kolmogorov–Smirnov (K-S) statistics of differences between the distributions estimated from the nonparametric dataset and the semiparametric/parametric datasets. For smokers (non-smokers), the K-S statistics for the semiparametric and parametric comparisons are 0.006 (0.002) and 0.05 (0.05), respectively. The semiparametric approach performs much better on the K-S metric.

**TABLE 1** Distribution of covariates by smoking status.

| | **Smoker** $n = 942$ | **Non-smoker** $n = 4054$ |
|---|---|---|
| Age (Mean/IQR) | 25.4/8.0 | 26.9/8.0 |
| 13–22 (%) | 32.5 | 23.6 |
| 23–27 (%) | 32.7 | 28.8 |
| 28–31 (%) | 20.9 | 26.3 |
| 31–45 (%) | 13.9 | 21.1 |
| Education (%) | | |
| Less than HS (%) | 32.5 | 14.6 |
| HS (%) | 50.8 | 43.3 |
| Greater than HS (%) | 16.7 | 42.1 |
| Race (%) | | |
| White (%) | 79.6 | 84.0 |
| Non-white (%) | 20.4 | 16.0 |
| Ethnicity (%) | | |
| Hispanic (%) | 3.0 | 4.0 |
| Non-hispanic (%) | 97.0 | 96.0 |
| Foreign (%) | | |
| Yes (%) | 2.7 | 6.2 |
| No (%) | 97.3 | 93.8 |
| Alcohol (%) | | |
| Yes (%) | 9.6 | 1.9 |
| No (%) | 90.4 | 98.1 |
| Marital status (%) | | |
| Married (%) | 47.3 | 74.6 |
| Not married (%) | 52.7 | 36.4 |
| Live birth order (%) | | |
| One | 34.3 | 43.0 |
| Two | 32.8 | 31.9 |
| Greater than two | 32.9 | 25.1 |
| Prenatal visits (Mean/IQR) | 9.7/5.0 | 10.8/4.0 |
| 0–9 (%) | 41.0 | 28.4 |
| 10–11 (%) | 22.4 | 23. |
| 12–13 (%) | 22.5 | 29.6 |
| 14–49 (%) | 14.1 | 18.2 |

In our analysis, we used $K = 5$ splits and a bi-weight kernel of the form: $\mathcal{K}(u) = \frac{15}{16}(1-u^2)^2 I(|u| \leq 1)$. Figure 2a displays the estimated means (solid lines) of $Y(1)$ and $Y(0)$ as a function of $\gamma_1$ and $\gamma_0$, respectively $(0.0 \leq \gamma_t \leq 0.0025)$. The figure includes pointwise 95% confidence intervals (dashed lines). The estimated effect of smoking on birth weight when $\gamma_1 = \gamma_0$ (ie, no unmeasured confounding) is −227 g (95% CI: −279 to −174).

To understand the choice of sensitivity parameters, consider Figure 2b. Here, we present the induced estimated mean of $Y(t)$ given $T = 1 - t$ as a function of $\gamma_t$. For fixed $\gamma_t$, the induced estimated mean is computed as $(\widehat{\psi}_t(\gamma_t) - E_n[Y|T = t]P_n[T = t])/P_n[T = 1 - t]$, where $E_n[Y|T = t]$ is the observed mean birth weight among individuals with $T = t$ and $P_n[T = t']$ is the observed proportion of individuals with smoking status $t'$. It makes clinical sense that $E[Y(1)|T = 0] \leq E[Y(0)|T = 0]$ and $E[Y(0)|T = 1] \geq E[Y(1)|T = 1]$. This restricts the value of $\gamma_0 \leq 0.0012$; there are no restrictions on $\gamma_1$.

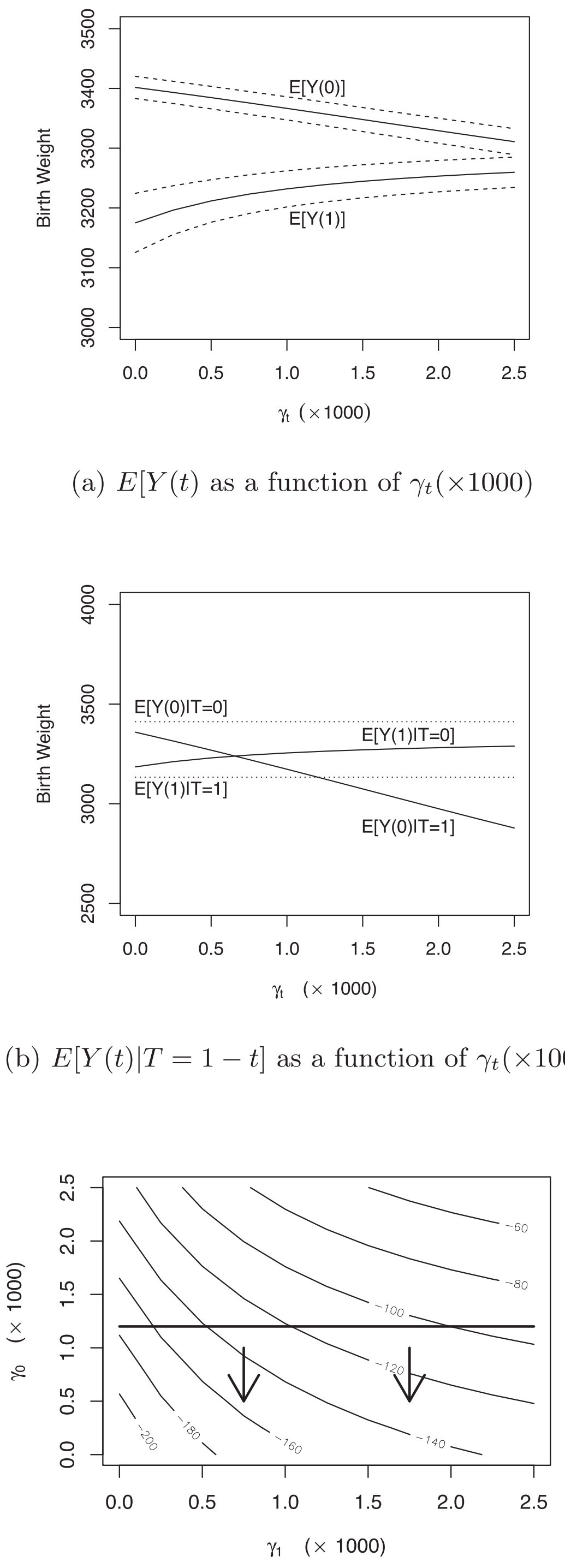

(a) $E[Y(t)$ as a function of $\gamma_t(\times 1000)$

(b) $E[Y(t)|T=1-t]$ as a function of $\gamma_t(\times 1000)$

(c) Contour plot of $E[Y(1)] - E[Y(0)]$ as a function of $\gamma_1(\times 1000)$ and $\gamma_0(\times 1000)$

**FIGURE 2** (a) Estimated means (solid lines) of $Y(t)$ and pointwise 95% confidence intervals (dashed lines) as a function of $\gamma_t$. (b) Indicated means of $E[Y(t)|T=1-t]$ as a function of $\gamma_t$. (c) Contour plot of estimated treatment effects as a function of $\gamma_1$ and $\gamma_0$; the region with arrows reflects the plausible range of sensitivity parameters.

**TABLE 2** True values.

| $\gamma_0/\gamma_1$ | | 0.00 | 0.25 | 0.50 | 0.75 | 1.00 | 1.25 | 1.50 | 1.75 | 2.00 | 2.25 | 2.50 |
|---|---|---|---|---|---|---|---|---|---|---|---|---|
| | | 3183.54$^{\dagger}$ | 3196.81 | 3208.96 | 3219.74 | 3229.12 | 3237.18 | 3244.07 | 3249.94 | 3254.96 | 3259.25 | 3262.95 |
| 0.00 | 3403.40* | −219.86$^{\bullet}$ | −210.21 | −200.47 | −190.69 | −180.93 | −171.24 | −161.67 | −152.28 | −143.14 | −134.29 | −125.79 |
| 0.25 | 3393.74 | −206.59 | −196.93 | −187.19 | −177.42 | −167.66 | −157.96 | −148.39 | −139.01 | −129.86 | −121.02 | −112.52 |
| 0.50 | 3384.01 | −194.44 | −184.78 | −175.04 | −165.27 | −155.50 | −145.81 | −136.24 | −126.85 | −117.71 | −108.86 | −100.37 |
| 0.75 | 3374.23 | −183.66 | −174.00 | −164.26 | −154.49 | −144.72 | −135.03 | −125.46 | −116.07 | −106.93 | −98.08 | −89.59 |
| 1.00 | 3364.47 | −174.28 | −164.62 | −154.89 | −145.11 | −135.35 | −125.65 | −116.08 | −106.7 | −97.55 | −88.71 | −80.21 |
| 1.25 | 3354.77 | −166.22 | −156.56 | −146.83 | −137.05 | −127.29 | −117.59 | −108.02 | −98.64 | −89.49 | −80.65 | −72.15 |
| 1.50 | 3345.20 | −159.33 | −149.68 | −139.94 | −130.16 | −120.40 | −110.71 | −101.14 | −91.75 | −82.61 | −73.76 | −65.26 |
| 1.75 | 3335.82 | −153.46 | −143.8 | −134.06 | −124.29 | −114.53 | −104.83 | −95.26 | −85.87 | −76.73 | −67.89 | −59.39 |
| 2.00 | 3326.67 | −148.44 | −138.79 | −129.05 | −119.27 | −109.51 | −99.81 | −90.24 | −80.86 | −71.72 | −62.87 | −54.37 |
| 2.25 | 3317.83 | −144.15 | −134.49 | −124.75 | −114.98 | −105.21 | −95.52 | −85.95 | −76.56 | −67.42 | −58.57 | −50.08 |
| 2.50 | 3309.33 | −140.45 | −130.8 | −121.06 | −111.28 | −101.52 | −91.83 | −82.26 | −72.87 | −63.73 | −54.88 | −46.38 |

$^{\dagger}E[Y(1)]$; $^{*}E[Y(0)]$; $^{\bullet}E[Y(1)] - E[Y(0)]$.

Figure 2c displays a contour plot of the ACE, $E[Y(1)] - E[Y(0)]$ as a function of the sensitivity parameters $\gamma_1$ and $\gamma_0$. For reference, we placed a horizontal line at $\gamma_0 = 0.0012$ and indicated the plausible region with arrows. Regardless of the choice of $\gamma_1$ and $\gamma_0$, the upper limit of the 95% confidence interval of the ACE is negative. This suggests a negative impact of maternal smoking on birth weight, which is consistent with the strong biological rationale reported in the literature (Albuquerque et al., 2004; Gozubuyuk et al., 2017).Our sensitivity analysis does suggest that the estimated ACE is likely smaller in magnitude than reported by Almond et al. (2005); it could be as small as an estimated reduction of 95 g.

In Supplementary Appendices D, E, and F, we present the results of three alternative approaches that have been considered in the literature: (1) Robins (1999), Brumback et al. (2004), Sjölander et al. (2022), and Lu and Ding (2023), (2) Cinelli and Hazlett (2020), and (3) Franks et al. (2020). While all approaches including ours indicate that smoking has a detrimental impact on birth weight; they differ with regard to the magnitude of the possible effects.

## 6 SIMULATION STUDY

To construct a realistic simulation study, we used the empirical distribution of $X$, the estimated distributions of $P(Y \leq y|T = t, X)$ and $P(T = 1|X)$ from the data analysis above as the true observed data generating mechanisms. We used the functional form for $q_t\{y, x; \gamma_t, h_t(F_t(\cdot|x); \gamma_t)\}$ discussed is Section 3.2. Using the observed data distribution and the choice for $q_t\{y, x; \gamma_t, h_t(F_t(\cdot|x); \gamma_t)\}$, we used (8) to compute the true value of $\mathbb{E}[Y(t)]$ as a function of $\gamma_t$ and the true values of $\mathbb{E}[Y(1) - Y(0)]$ as a function of $\gamma_1$ and $\gamma_0$ (see Table 2). In our simulation, we considered sample sizes of 1000, 2500, and 5000. For each sample size, we simulated 2000 datasets. For each simulated dataset, we used our estimation procedure to estimate $\mathbb{E}[Y(t)]$ as a function of $\gamma_t$ and $\mathbb{E}[Y(1) - Y(0)]$ as a function of $\gamma_1$ and $\gamma_0$. We evaluated estimation bias and 95% confidence interval coverage. In Table 3, we report on bias and 95% confidence interval coverage. Bias is low for all choices of sensitivity parameters and reduces with sample size. Confidence interval coverage is close to the nominal level for all sample sizes and choice of sensitivity parameters.

Supplementary Appendix G presents the results of a simulation study where the model generating the observed data is different than the GAM/single-index models used for analysis. Bias is low and confidence interval coverage is close to the nominal level.

## 7 DISCUSSION

We developed a semiparametric method for assessing sensitivity to the "no unmeasured confounding" assumption, which when true, implies that the ACE is identified via covariate adjustment. We generalized the sensitivity analysis approach developed by Robins et al. (2000), Franks et al. (2020), and Zhou and Yao (2023). Specifically, we derived the influence function associated with a broader class of tilting functions than considered in these papers. This broader class involves more sensitivity parameters which can make the analysis more complex and impact estimates in non-trivial ways. However, as illustrated by our maternal smoking–birth weight example, introducing complexity may be required to match the clinical judgement that is unavoidable when dealing with unmeasured confounding.

The adjustment functional (8) is not the only way the ACE can be identified. For instance, in the "front door" model where there is measured mediator with no unmeasured causes that captures all the effect of the treatment on outcome, the ACE is identifiable via a more complicated functional (Pearl, 2009). There exist more examples of identification despite the presence of unmeasured confounders; see (Shpitser and Pearl, 2006; Bhattacharya et al., 2022). The semiparametric sensitivity analysis procedure described here can be adapted to assess the underlying assumptions in those models in a robust way. This opens up several interesting directions for future work.

In the interest of simplicity and concreteness, we chose to use GAMs and single index models to estimate the nuisance functions. Due to our use of sample splitting and influence function-based estimators, all the asymptotic results will continue to hold for any nuisance estimators that converge to the truth faster than $n^{-1/4}$ rates. For example, we could have used sufficient dimension reduction techniques (see, eg, Ma and Zhu [2012, 2013]; Huang and Chiang [2017]). Alternatively, we could have used a black-box ensemble method such as Super Learner (van der Laan et al., 2007); inference would then still be asymptotically

**TABLE 3** Results of simulation study for various sample sizes.

| $\gamma_0/\gamma_1$ (×1000) | *n* | | **0.00** | **0.25** | **0.50** | **0.75** | **1.00** | **1.25** | **1.50** | **1.75** | **2.00** | **2.25** | **2.50** |
|---|---|---|---|---|---|---|---|---|---|---|---|---|---|
| | 1000 | | −16.96† | −14.49 | −13.14 | −12.36 | −11.90 | −11.58 | −11.29 | −11.16 | −11.10 | −11.07 | −11.10 |
| | | | 0.94† | *0.94* | *0.94* | *0.95* | *0.95* | *0.95* | *0.94* | *0.94* | *0.94* | *0.94* | *0.94* |
| | 2500 | | −4.46 | −3.66 | −3.20 | −2.97 | −2.88 | −2.90 | −2.98 | −3.07 | −3.17 | −3.25 | −3.35 |
| | | | *0.95* | *0.95* | *0.95* | *0.95* | *0.95* | *0.95* | *0.95* | *0.95* | *0.95* | *0.95* | *0.95* |
| | 5000 | | −1.73 | −1.64 | −1.61 | −1.65 | −1.71 | −1.78 | −1.87 | −1.94 | −2.01 | −2.09 | −2.16 |
| | | | *0.95* | *0.95* | *0.95* | *0.95* | *0.95* | *0.95* | *0.95* | *0.95* | *0.94* | *0.94* | *0.94* |
| 0.00 | 1000 | −0.48* | −16.48• | −16.16 | −15.82 | −15.49 | −15.16 | −14.82 | −14.50 | −14.18 | −13.87 | −13.59 | −13.33 |
| | | 0.95* | *0.94* | *0.95* | *0.95* | *0.95* | *0.95* | *0.95* | *0.95* | *0.95* | *0.95* | *0.95* | *0.95* |
| | 2500 | 0.25 | −4.71 | −4.55 | −4.38 | −4.21 | −4.04 | −3.88 | −3.72 | −3.57 | −3.42 | −3.29 | −3.16 |
| | | *0.96* | *0.95* | *0.95* | *0.96* | *0.96* | *0.96* | *0.95* | *0.95* | *0.95* | *0.95* | *0.95* | *0.95* |
| | 5000 | −0.27 | −1.46 | −1.37 | −1.28 | −1.20 | −1.11 | −1.03 | −0.95 | −0.88 | −0.81 | −0.74 | −0.68 |
| | | *0.95* | *0.95* | *0.96* | *0.96* | *0.96* | *0.96* | *0.96* | *0.95* | *0.96* | *0.96* | *0.96* | *0.96* |
| 0.25 | 1000 | −0.80 | −14.02 | −13.69 | −13.36 | −13.03 | −12.69 | −12.36 | −12.04 | −11.72 | −11.41 | −11.13 | −10.87 |
| | | *0.95* | *0.95* | *0.95* | *0.95* | *0.95* | *0.95* | *0.95* | *0.95* | *0.95* | *0.95* | *0.95* | *0.95* |
| | 2500 | 0.09 | −3.92 | −3.75 | −3.58 | −3.41 | −3.25 | −3.08 | −2.92 | −2.77 | −2.63 | −2.50 | −2.37 |
| | | *0.96* | *0.95* | *0.95* | *0.95* | *0.96* | *0.96* | *0.96* | *0.96* | *0.95* | *0.96* | *0.96* | *0.95* |
| | 5000 | −0.36 | −1.36 | −1.28 | −1.19 | −1.10 | −1.02 | −0.94 | −0.86 | −0.79 | −0.72 | −0.65 | −0.59 |
| | | *0.95* | *0.95* | *0.95* | *0.95* | *0.95* | *0.95* | *0.95* | *0.95* | *0.95* | *0.95* | *0.96* | *0.95* |
| 0.50 | 1000 | −1.13 | −12.67 | −12.34 | −12.01 | −11.67 | −11.34 | −11.01 | −10.68 | −10.36 | −10.06 | −9.77 | −9.51 |
| | | *0.96* | *0.95* | *0.95* | *0.95* | *0.95* | *0.95* | *0.95* | *0.95* | *0.95* | *0.95* | *0.95* | *0.95* |
| | 2500 | −0.08 | −3.45 | −3.28 | −3.12 | −2.95 | −2.78 | −2.62 | −2.46 | −2.31 | −2.16 | −2.03 | −1.90 |
| | | *0.96* | *0.95* | *0.95* | *0.95* | *0.96* | *0.96* | *0.95* | *0.95* | *0.95* | *0.96* | *0.96* | *0.96* |
| | 5000 | −0.45 | −1.34 | −1.26 | −1.17 | −1.08 | −1.00 | −0.92 | −0.84 | −0.76 | −0.69 | −0.63 | −0.57 |
| | | *0.95* | *0.95* | *0.95* | *0.95* | *0.95* | *0.95* | *0.95* | *0.95* | *0.95* | *0.95* | *0.95* | *0.95* |
| 0.75 | 1000 | −1.47 | −11.88 | −11.56 | −11.23 | −10.89 | −10.56 | −10.23 | −9.90 | −9.58 | −9.27 | −8.99 | −8.73 |
| | | *0.95* | *0.95* | *0.95* | *0.95* | *0.95* | *0.95* | *0.95* | *0.95* | *0.95* | *0.95* | *0.95* | *0.95* |
| | 2500 | −0.25 | −3.22 | −3.06 | −2.89 | −2.72 | −2.55 | −2.39 | −2.23 | −2.08 | −1.93 | −1.80 | −1.67 |
| | | *0.96* | *0.96* | *0.96* | *0.96* | *0.96* | *0.96* | *0.96* | *0.96* | *0.96* | *0.96* | *0.96* | *0.96* |
| | 5000 | −0.53 | −1.38 | −1.29 | −1.20 | −1.11 | −1.03 | −0.95 | −0.87 | −0.80 | −0.73 | −0.66 | −0.60 |
| | | *0.95* | *0.95* | *0.96* | *0.96* | *0.95* | *0.96* | *0.96* | *0.95* | *0.96* | *0.96* | *0.96* | *0.96* |
| 1.00 | 1000 | −1.80 | −11.43 | −11.10 | −10.77 | −10.44 | −10.10 | −9.77 | −9.45 | −9.13 | −8.82 | −8.54 | −8.28 |
| | | *0.96* | *0.95* | *0.95* | *0.95* | *0.95* | *0.95* | *0.95* | *0.95* | *0.95* | *0.95* | *0.95* | *0.95* |
| | 2500 | −0.42 | −3.14 | −2.97 | −2.80 | −2.63 | −2.46 | −2.30 | −2.14 | −1.99 | −1.85 | −1.71 | −1.59 |
| | | *0.96* | *0.96* | *0.96* | *0.96* | *0.96* | *0.96* | *0.96* | *0.96* | *0.96* | *0.96* | *0.96* | *0.96* |
| | 5000 | -0.62 | −1.44 | −1.35 | −1.26 | −1.18 | −1.09 | −1.01 | −0.93 | −0.86 | −0.79 | −0.72 | −0.66 |
| | | *0.95* | *0.96* | *0.96* | *0.96* | *0.96* | *0.96* | *0.96* | *0.96* | *0.96* | *0.96* | *0.96* | *0.96* |
| 1.25 | 1000 | −2.13 | −11.10 | −10.78 | −10.44 | −10.11 | −9.78 | −9.44 | −9.12 | −8.80 | −8.49 | −8.21 | −7.95 |
| | | *0.96* | *0.96* | *0.96* | *0.96* | *0.95* | *0.96* | *0.96* | *0.96* | *0.95* | *0.95* | *0.95* | *0.95* |
| | 2500 | −0.58 | −3.16 | −2.99 | −2.82 | −2.65 | −2.49 | −2.32 | −2.16 | −2.01 | −1.87 | −1.73 | −1.61 |
| | | *0.96* | *0.96* | *0.96* | *0.96* | *0.96* | *0.96* | *0.96* | *0.96* | *0.96* | *0.96* | *0.96* | *0.96* |
| | 5000 | −0.70 | −1.51 | −1.43 | −1.34 | −1.25 | −1.17 | −1.09 | −1.01 | −0.93 | −0.86 | −0.80 | −0.74 |
| | | *0.95* | *0.95* | *0.95* | *0.95* | *0.95* | *0.96* | *0.96* | *0.95* | *0.96* | *0.96* | *0.95* | *0.95* |
| 1.50 | 1000 | −2.46 | −10.82 | −10.49 | −10.16 | −9.83 | −9.49 | −9.16 | −8.84 | −8.52 | −8.21 | −7.93 | −7.67 |
| | | *0.96* | *0.95* | *0.95* | *0.95* | *0.95* | *0.96* | *0.95* | *0.95* | *0.95* | *0.95* | *0.95* | *0.95* |
| | 2500 | −0.74 | −3.23 | −3.06 | −2.89 | −2.73 | −2.56 | −2.39 | −2.23 | −2.08 | −1.94 | −1.81 | −1.68 |
| | | *0.96* | *0.96* | *0.96* | *0.96* | *0.96* | *0.96* | *0.96* | *0.96* | *0.96* | *0.96* | *0.96* | *0.96* |
| | 5000 | −0.78 | −1.60 | −1.51 | −1.42 | −1.33 | −1.25 | −1.17 | −1.09 | −1.02 | −0.95 | −0.88 | −0.82 |
| | | *0.95* | *0.95* | *0.95* | *0.95* | *0.95* | *0.95* | *0.95* | *0.95* | *0.95* | *0.95* | *0.95* | *0.95* |
| 1.75 | 1000 | -2.78 | −10.68 | −10.36 | −10.02 | −9.69 | −9.36 | −9.02 | −8.70 | −8.38 | −8.07 | −7.79 | −7.53 |
| | | *0.96* | *0.95* | *0.95* | *0.95* | *0.95* | *0.95* | *0.95* | *0.95* | *0.95* | *0.95* | *0.95* | *0.95* |
| | 2500 | −0.89 | −3.32 | −3.16 | −2.99 | −2.82 | −2.65 | −2.49 | −2.33 | −2.18 | −2.04 | −1.90 | −1.77 |
| | | *0.96* | *0.96* | *0.96* | *0.96* | *0.96* | *0.96* | *0.96* | *0.96* | *0.96* | *0.96* | *0.96* | *0.96* |
| | 5000 | −0.85 | −1.67 | −1.58 | −1.50 | −1.41 | −1.33 | −1.24 | −1.17 | −1.09 | −1.02 | −0.96 | −0.90 |
| | | *0.96* | *0.95* | *0.95* | *0.95* | *0.95* | *0.95* | *0.95* | *0.95* | *0.95* | *0.95* | *0.95* | *0.95* |
| 2.00 | 1000 | −3.09 | −10.62 | −10.30 | −9.97 | −9.63 | −9.30 | −8.97 | −8.64 | −8.32 | −8.01 | −7.73 | −7.47 |
| | | *0.96* | *0.95* | *0.95* | *0.95* | *0.95* | *0.95* | *0.95* | *0.95* | *0.95* | *0.95* | *0.95* | *0.95* |
| | 2500 | −1.04 | −3.42 | −3.26 | −3.09 | −2.92 | −2.75 | −2.59 | −2.43 | −2.28 | −2.13 | −2.00 | −1.87 |
| | | *0.96* | *0.96* | *0.96* | *0.96* | *0.96* | *0.96* | *0.96* | *0.96* | *0.96* | *0.96* | *0.96* | *0.96* |
| | 5000 | −0.92 | −1.74 | −1.66 | −1.57 | −1.48 | −1.40 | −1.32 | −1.24 | −1.16 | −1.09 | −1.03 | −0.97 |
| | | *0.96* | *0.95* | *0.95* | *0.95* | *0.95* | *0.95* | *0.95* | *0.95* | *0.95* | *0.95* | *0.96* | *0.95* |

**TABLE 3** Continued

| $\gamma_0/\gamma_1$ ($\times 1000$) | $n$ | | 0.00 | 0.25 | 0.50 | 0.75 | 1.00 | 1.25 | 1.50 | 1.75 | 2.00 | 2.25 | 2.50 |
|---|---|---|---|---|---|---|---|---|---|---|---|---|---|
| 2.25 | 1000 | −3.37 | −10.60 | −10.27 | −9.94 | −9.60 | −9.27 | −8.94 | −8.61 | −8.29 | −7.99 | −7.70 | −7.44 |
| | | *0.96* | *0.95* | *0.95* | *0.95* | *0.95* | *0.95* | *0.95* | *0.95* | *0.95* | *0.95* | *0.95* | *0.95* |
| | 2500 | −1.17 | −3.51 | −3.34 | −3.17 | −3.00 | −2.84 | −2.67 | −2.51 | −2.36 | −2.22 | −2.09 | −1.96 |
| | | *0.96* | *0.95* | *0.95* | *0.95* | *0.95* | *0.95* | *0.96* | *0.96* | *0.96* | *0.96* | *0.96* | *0.96* |
| | 5000 | -0.99 | −1.81 | −1.73 | −1.64 | −1.55 | −1.47 | −1.39 | −1.31 | −1.23 | −1.16 | −1.10 | −1.04 |
| | | *0.95* | *0.95* | *0.95* | *0.95* | *0.95* | *0.95* | *0.95* | *0.95* | *0.95* | *0.96* | *0.96* | *0.95* |
| 2.50 | 1000 | −3.63 | −10.63 | −10.30 | −9.97 | −9.63 | −9.30 | −8.97 | −8.64 | −8.32 | −8.02 | −7.73 | −7.47 |
| | | *0.96* | *0.95* | *0.95* | *0.95* | *0.95* | *0.95* | *0.95* | *0.95* | *0.95* | *0.95* | *0.95* | *0.95* |
| | 2500 | −1.30 | −3.60 | −3.44 | −3.27 | −3.10 | −2.93 | −2.77 | −2.61 | −2.46 | −2.32 | −2.18 | −2.05 |
| | | *0.96* | *0.95* | *0.95* | *0.95* | *0.96* | *0.96* | *0.96* | *0.96* | *0.96* | *0.96* | *0.96* | *0.96* |
| | 5000 | −1.05 | −1.89 | −1.80 | −1.71 | −1.62 | −1.54 | −1.46 | −1.38 | −1.31 | −1.24 | −1.17 | −1.11 |
| | | *0.95* | *0.95* | *0.95* | *0.95* | *0.95* | *0.95* | *0.95* | *0.95* | *0.95* | *0.95* | *0.95* | *0.95* |

Regular font denotes bias. Italics denotes confidence interval coverage. $^{\dagger}E[Y(1)]$; $^{*}E[Y(0)]$; $^{\bullet}E[Y(1)] - E[Y(0)]$.

valid, as long as at least one learner has root mean squared error that scales faster than $n^{-1/4}$. Achieving root mean square error of smaller order than $n^{-1/4}$ requires that the nuisance functions belong in a function class that is not too complex. Under certain conditions, this requirement could be weakened by incorporating higher order influence function terms that would estimate the remainder $\text{Rem}_t(\widehat{P}, P; \gamma_t)$ (Robins et al., 2008, 2017). This results in estimators that are considerably more complex than the (first-order) influence-function based estimator that we have constructed, and is thus beyond the scope of this paper.

## SUPPLEMENTARY MATERIALS

Supplementary material is available at *Biometrics* online.

Web Appendices and data/code referenced in Sections 4, 5, and 6 are available with this paper at the Biometrics website on Oxford Academic.

## FUNDING

None declared.

## CONFLICT OF INTEREST

None declared.

## DATA AVAILABILITY

The birth weight dataset used in this article is available in the online supplementary material.

## REFERENCES

Albuquerque, C. A., Smith, K. R., Johnson, C., Chao, R. and Harding, R. (2004). Influence of maternal tobacco smoking during pregnancy on uterine, umbilical and fetal cerebral artery blood flows. *Early Human Development*, 80, 31–42.

Almond, D., Chay, K. Y. and Lee, D. S. (2005). The costs of low birth weight. *The Quarterly Journal of Economics*, 120, 1031–1083.

Berard, J., Dufour, P., Vinatier, D., Subtil, D., Vanderstichele, S., Monnier, J. et al. (1998). Fetal macrosomia: risk factors and outcome: A study of the outcome concerning 100 cases >4500 g. *European Journal of Obstetrics & Gynecology and Reproductive Biology*, 77, 51–59.

Bhattacharya, R., Nabi, R. and Shpitser, I. (2022). Semiparametric inference for causal effects in graphical models with hidden variables. *The Journal of Machine Learning Research*, 23, 13325–13400.

Bonvini, M. and Kennedy, E. H. (2022). Sensitivity analysis via the proportion of unmeasured confounding. *Journal of the American Statistical Association*, 117, 1540–1550.

Brumback, B. A., Hernán, M. A., Haneuse, S. J. and Robins, J. M. (2004). Sensitivity analyses for unmeasured confounding assuming a marginal structural model for repeated measures. *Statistics in Medicine*, 23, 749–767.

Carnegie, N. B., Harada, M. and Hill, J. L. (2016). Assessing sensitivity to unmeasured confounding using a simulated potential confounder. *Journal of Research on Educational*, 9, 395–420.

Chiang, C.-T. and Huang, M.-Y. (2012). New estimation and inference procedures for a single-index conditional distribution model. *Journal of Multivariate Analysis*, 111, 271–285.

Cinelli, C. and Hazlett, C. (2020). Making sense of sensitivity: Extending omitted variable bias. *Journal of the Royal Statistical Society: Series B*, 82, 39–67.

Cornfield, J., Haenszel, W., Hammond, E. C., Lilienfeld, A. M., Shimkin, M. B. and Wynder, E. L. (1959). Smoking and lung cancer: recent evidence and a discussion of some questions. *Journal of the National Cancer Institute*, 22, 173–203.

Díaz, I., Luedtke, A. R. and van der Laan, M. J. (2018). Sensitivity analysis. In: *Targeted Learning in Data Science*, 511–522. New York, USA: Springer.

Díaz, I. and van der Laan, M. J. (2013). Sensitivity analysis for causal inference under unmeasured confounding and measurement error problems. *The International Journal of Biostatistics*, 9, 149–160.

Ding, P. and VanderWeele, T. J. (2016). Sensitivity analysis without assumptions. *Epidemiology*, 27, 368.

Dorie, V., Harada, M., Carnegie, N. B. and Hill, J. (2016). A flexible, interpretable framework for assessing sensitivity to unmeasured confounding. *Statistics in Medicine*, 35, 3453–3470.

Dorn, J. and Guo, K. (2022). Sharp sensitivity analysis for inverse propensity weighting via quantile balancing. *Journal of the American Statistical Association*, 118, 1–13.

Dorn, J., Guo, K. and Kallus, N. (2024). Doubly-valid/doubly-sharp sensitivity analysis for causal inference with unmeasured confounding. *Journal of the American Statistical Association*, 1–12.

Franks, A., D'Amour, A. and Feller, A. (2020). Flexible sensitivity analysis for observational studies without observable implications. *Journal of the American Statistical Association*, 115, 1730–1746.

Gastwirth, J. L., Krieger, A. M. and Rosenbaum, P. R. (1998). Dual and simultaneous sensitivity analysis for matched pairs. *Biometrika*, 85, 907–920.

Gozubuyuk, A. A., Dag, H., Kaçar, A., Karakurt, Y. and Arica, V. (2017). Epidemiology, pathophysiology, clinical evaluation, and treatment of carbon monoxide poisoning in child, infant, and fetus. *Northern Clinics of Istanbul*, 4, 100–107.

Hastie, T. J. and Tibshirani, R. J. (2017). *Generalized Additive Models*. New York, USA: Routledge.

Hernán, M. A. and Robins, J. M. (2006). Estimating causal effects from epidemiological data. *Journal of Epidemiology & Community Health*, 60, 578–586.

Horowitz, J. L. and Mammen, E. (2004). Nonparametric estimation of an additive model with a link function. *The Annals of Statistics*, 32, 2412–2443.

Huang, M.-Y. and Chiang, C.-T. (2017). An effective semiparametric estimation approach for the sufficient dimension reduction model. *Journal of the American Statistical Association*, 112, 1296–1310.

Imbens, G. W. (2003). Sensitivity to exogeneity assumptions in program evaluation. *American Economic Review*, 93, 126–132.

Kennedy, E. H.. (2022). Semiparametric doubly robust targeted double machine learning: a review. preprint (arXiv:2203.06469).

Lu, S. and Ding, P. (2023). Flexible sensitivity analysis for causal inference in observational studies subject to unmeasured confounding. preprint (arXiv:2305.17643).

Ma, Y. and Zhu, L. (2012). A semiparametric approach to dimension reduction. *Journal of the American Statistical Association*, 107, 168–179.

Ma, Y. and Zhu, L. (2013). Efficient estimation in sufficient dimension reduction. *The Annals of Statistics*, 41, 250.

Manski, C. F. (1990). Nonparametric bounds on treatment effects. *The American Economic Review*, 80, 319–323.

Neyman, J. (1923). Sur les applications de la thar des probabilities aux experiences agaricales: Essay des principle. excerpts reprinted (1990) in English. *Statistical Science*, 5, 463–472.

Palmsten, K., Nelson, K. K., Laurent, L. C., Park, S., Chambers, C. D. and Parast, M. M. (2018). Subclinical and clinical chorioamnionitis, fetal vasculitis, and risk for preterm birth: A cohort study. *Placenta*, 67, 54–60.

Pearl, J. (2009). *Causality: Models, Reasoning, and Inference*. Cambridge, United Kingdom: Cambridge University Press.

Robins, J., Li, L., Tchetgen, E. and van der Vaart, A. (2008). Higher order influence functions and minimax estimation of nonlinear functionals. In: *Probability and Statistics: Essays in Honor of David A. Freedman*, 335–421. Beachwood, Ohio, USA: Institute of Mathematical Statistics.

Robins, J. M. (1986). A new approach to causal inference in mortality studies with sustained exposure periods–application to control of the healthy worker survivor effect. *Mathematical Modeling*, 7, 1393–1512.

Robins, J. M. (1989). The analysis of randomized and non-randomized aids treatment trials using a new approach to causal inference in longitudinal studies. In: *Health Service Research Methodology: A Focus on AIDS*, 113–159. Rockville, MD: US Department of Healthand Human Services.

Robins, J. M. (1999). Association, causation, and marginal structural models. *Synthese*, 121, 151–179.

Robins, J. M., Li, L., Mukherjee, R., Tchetgen, E. T. and van der Vaart, A. (2017). Minimax estimation of a functional on a structured high-dimensional model. *Annals of Statistics*, 45, 1951–1987.

Robins, J. M., Rotnitzky, A. and Scharfstein, D. O. (2000). Sensitivity analysis for selection bias and unmeasured confounding in missing data and causal inference models. In: *Statistical Models in Epidemiology, the Environment, and Clinical Trials*, 1–94. New York, USA: Springer.

Robins, J. M., Rotnitzky, A. and Zhao, L. P. (1994). Estimation of regression coefficients when some regressors are not always observed. *Journal of the American Statistical Association*, 89, 846–866.

Rosenbaum, P. R. (1987). Sensitivity analysis for certain permutation inferences in matched observational studies. *Biometrika*, 74, 13–26.

Rosenbaum, P. R. and Rubin, D. B. (1983a). Assessing sensitivity to an unobserved binary covariate in an observational study with binary outcome. *Journal of the Royal Statistical Society: Series B*, 45, 212–218.

Rosenbaum, P. R. and Rubin, D. B. (1983b). The central role of the propensity score in observational studies for causal effects. *Biometrika*, 70, 41–55.

Rotnitzky, A., Smucler, E. and Robins, J. M. (2021). Characterization of parameters with a mixed bias property. *Biometrika*, 108, 231–238.

Rubin, D. B. (1974). Estimating causal effects of treatments in randomized and non-randomized studies. *Journal of Educational Psychology*, 66, 688–701.

Salafia, C. M., Ernst, L. M., Pezzullo, J. C., Wolf, E. J., Rosenkrantz, T. S. and Vintzileos, A. M. (1995). The very low birthweight infant: maternal complications leading to preterm birth, placental lesions, and intrauterine growth. *American Journal of Perinatology*, 12, 106–110.

Salafia, C. M., Minior, V. K., Pezzullo, J. C., Popek, E. J., Rosenkrantz, T. S. and Vintzileos, A. M. (1995). Intrauterine growth restriction in infants of less than thirty-two weeks' gestation: associated placental pathologic features. *American Journal of Obstetrics and Gynecology*, 173, 1049–1057.

Scharfstein, D. O. and Irizarry, R. A. (2003). Generalized additive selection models for the analysis of studies with potentially nonignorable missing outcome data. *Biometrics*, 59, 601–613.

Scharfstein, D. O., Rotnitzky, A. and Robins, J. M. (1999). Adjusting for nonignorable drop-out using semiparametric nonresponse models. *Journal of the American Statistical Association*, 94, 1096–1120.

Shen, C., Li, X., Li, L. and Were, M. C. (2011). Sensitivity analysis for causal inference using inverse probability weighting. *Biometrical Journal*, 53, 822–837.

Shpitser, I. and Pearl, J. (2006). Identification of joint interventional distributions in recursive semi-Markovian causal models. In: *Proceedings of the Twenty-First National Conference on Artificial Intelligence (AAAI-06)*. Palo Alto, CA: AAAI Press.

Sjölander, A., Gabriel, E. E. and Ciocănea-Teodorescu, I. (2022). Sensitivity analysis for causal effects with generalized linear models. *Journal of Causal Inference*, 10, 441–479.

Stotland, N., Caughey, A., Breed, E. and Escobar, G. (2004). Risk factors and obstetric complications associated with macrosomia. *International Journal of Gynecology & Obstetrics*, 87, 220–226.

Tan, Z. (2006). A distributional approach for causal inference using propensity scores. *Journal of the American Statistical Association*, 101, 1619–1637.

Tsiatis, A. A. (2006). *Semiparametric Theory and Missing Data*. New York, USA: Springer.

van der Laan, M. J., Polley, E. C. and Hubbard, A. E. (2007). Super learner. *Statistical Applications in Genetics and Molecular Biology*, 6, Article 25.

van Der Laan, M. J. and Rubin, D. (2006). Targeted maximum likelihood learning. *The International Journal of Biostatistics*, 2, 11.

VanderWeele, T. J. and Arah, O. A. (2011). Bias formulas for sensitivity analysis of unmeasured confounding for general outcomes, treatments, and confounders. *Epidemiology*, 22, 42–52.

Veitch, V. and Zaveri, A. (2020). Sense and sensitivity analysis: Simple post-hoc analysis of bias due to unobserved confounding. *Advances in Neural Information Processing Systems*, 33, 10999–11009.

Wang, L., Zheng, C., Zhou, W. and Zhou, W.-X. (2021). A new principle for tuning-free huber regression. *Statistica Sinica*, 31, 2153–2177.

Yadlowsky, S., Namkoong, H., Basu, S., Duchi, J. and Tian, L. (2018). Bounds on the conditional and average treatment effect with unobserved confounding factors. *Annals of statistics*, 50, 2587–2615.

Zhang, B. and Tchetgen Tchetgen, E. J. (2022). A semi-parametric approach to model-based sensitivity analysis in observational studies. *Journal of the Royal Statistical Society Series A*, 185, S668–S691.

Zhao, Q., Small, D. S. and Bhattacharya, B. B. (2019). Sensitivity analysis for inverse probability weighting estimators via the percentile bootstrap. *Journal of the Royal Statistical Society: Series B*, 81, 735–761.

Zhou, M. and Yao, W. (2023). Sensitivity analysis of unmeasured confounding in causal inference based on exponential tilting and super learner. *Journal of Applied Statistics*, 50, 744–760.